\documentclass[a4paper, 9pt]{article}

\usepackage{INTERSPEECH2020}

\usepackage{caption} 
\usepackage{multirow} 
\usepackage{hyperref}

\usepackage{subcaption}

\title{A Preliminary Study of a Two-Stage Paradigm for Preserving Speaker Identity in Dysarthric Voice Conversion}
\name{\begin{tabular}{c}
	Wen-Chin Huang$^1$, Kazuhiro Kobayashi$^1$, Yu-Huai Peng$^2$, Ching-Feng Liu$^3$,\\
	Yu Tsao$^2$, Hsin-Min Wang$^2$, Tomoki Toda$^1$
	\end{tabular}
}
\address{$^1$Nagoya University, Japan
		$^2$Academia Sinica, Taiwan
		$^3$Chi Mei Hospital}
\email{wen.chinhuang@g.sp.m.is.nagoya-u.ac.jp}

\begin{document}

\maketitle
\begin{abstract}
We propose a new paradigm for maintaining speaker identity in dysarthric voice conversion (DVC). The poor quality of dysarthric speech can be greatly improved by statistical VC, but as the normal speech utterances of a dysarthria patient are nearly impossible to collect, previous work failed to recover the individuality of the patient. In light of this, we suggest a novel, two-stage approach for DVC, which is highly flexible in that no normal speech of the patient is required. First, a powerful parallel sequence-to-sequence model converts the input dysarthric speech into a normal speech of a reference speaker as an intermediate product, and a nonparallel, frame-wise VC model realized with a variational autoencoder then converts the speaker identity of the reference speech back to that of the patient while assumed to be capable of preserving the enhanced quality. We investigate several design options. Experimental evaluation results demonstrate the potential of our approach to improving the quality of the dysarthric speech while maintaining the speaker identity.
\end{abstract}

\vspace*{-3mm}
\section{Introduction}
\label{sec:intro}

Dysarthria refers to a type of speech disorder caused by disruptions in the neuromotor interface such as cerebral palsy or amyotrophic lateral sclerosis \cite{speech-motor-review}. Dysarthria patients lack normal control of the primary vocal articulators, resulting in abnormal and unintelligible speech with phoneme loss, unstable prosody, and imprecise articulation. The ability to communicate with speech in everyday life is therefore degraded, and it is of urgent need to improve the intelligibility of the distorted dysarthric speech.\footnote{In the field of VC, orthogonal descriptions such as ``naturalness'' and ``intelligibility'' are often used, but we use the term ``quality'' in this paper interchangeably.}

Voice conversion (VC), a technique that aims to convert the speech from a source to that of a target without changing the linguistic content \cite{VC}, has been a dominant approach for dysarthric speech enhancement. We hereafter refer to this task as DVC. Rule-based transformation based on signal processing \cite{dysarthric-adjust} is limited in that each patient needs to be individually considered. Statistical approaches adopt models ranging from Gaussian mixture models \cite{dysarthric-intelligibility}, exemplar-based methods \cite{dysarthric-nmf, dysarthric-hybrid} and deep neural networks \cite{dysarthric-gated-cnn, dysarthric-e2e, dysarthric-discogan}. 

One of the most difficult problems in not only DVC but VC for other disordered speech such as alaryngeal speech \cite{alaryngeal-improve} is how to maintain the patient identity after conversion. This is mainly because collecting normal speech of the patient is almost impossible.
There have been attempts to tackle this problem. A one-to-many VC system based on eigenvoice conversion was proposed for alaryngeal speech enhancement, whose setting was still considered too idealized since they assumed that a few normal samples of the patient can still be accessed \cite{alaryngeal-evc}.

\begin{figure*}[t]
	\centering
	\includegraphics[width=\textwidth]{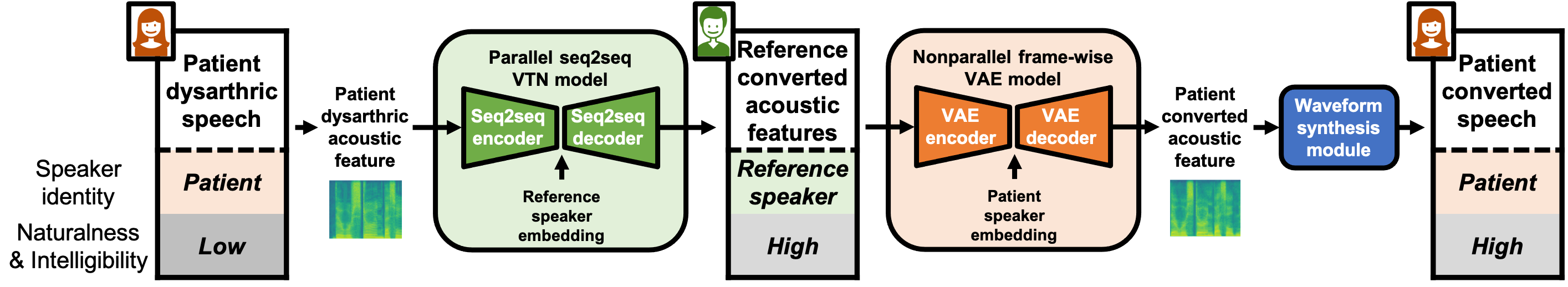}
	\centering
	\captionof{figure}{Illustration of the conversion process in the proposed two-stage method for preserving speaker identity in dysarthric voice conversion.}
	\label{fig:method}
	\vspace*{-3mm}
\end{figure*}

Our goal in this work is to utilize VC techniques to convert the patient's dysarthric speech into a more intelligible, more natural speech while maintaining the speaker identity of the patient. In light of this, we propose a novel, two-stage approach that combines recent advances in the field of VC. Figure~\ref{fig:method} depicts the general idea of the proposed method. In the first stage, a sequence-to-sequence (seq2seq) model converts the input dysarthric speech into that of a reference normal speaker, where we adopted a Transformer-based model named Voice Transformer Network (VTN) \cite{VTN}. The ability of seq2seq VC models to convert suprasegmental information and the parallel training strategy can greatly improve the naturalness and intelligibility, though the speaker identity is changed into that of the reference speaker. Next, a frame-wise, nonparallel VC model realized by a variational autoencoder (VAE) \cite{VAE, VAE-VC, crank} takes the converted speech with the identity of the reference speaker as input and restores the identity of the patient. An important assumption we make here is that due to the frame-wise constraint, the VAE model changes only time-invariant characteristics such as the speaker identity, while preserving time-variant characteristics, such as pronunciation. As a result, the converted speech has the speaker identity of the patient while maintaining high intelligibility and naturalness. We acknowledge that recently a very similar idea was proposed for preserving speaker identity in not DVC but dysarthric TTS \cite{dysarthric-tts-cyclevae}.

We evaluate our proposed method on a Mandarin corpus collected from a female cerebral palsy patient. We investigate the importance of the reference speaker choice, and examine how much the aforementioned assumption holds with the current VAE model we adopt. Finally, objective and subjective evaluations show that our approach can improve the naturalness and intelligibility of the dysarthric speech.

Our main contributions in this work are as follows:
\begin{itemize}
	\item We show that the proposed two-stage method for DVC can restore the patient identity without any normal speech of the patient while improving naturalness and intelligibility.
	\item To our knowledge, this is the first work to evaluate seq2seq modeling for DVC with a more complex dataset rather than single-worded datasets \cite{dysarthric-e2e}.
\end{itemize}

\section{Related works}

\subsection{Sequence-to-sequence voice conversion}

Compared with conventional frame-wise VC methods \cite{VC, GMM-VC}, seq2seq VC has shown its extraordinary ability in converting speaker identity \cite{VTN, ATT-S2S-VC, S2S-iFLYTEK-VC, S2S-Text-VC, S2S-parrotron-VC}. Seq2seq modeling is capable of generating outputs of various lengths and capturing long-term dependencies \cite{S2S}, making it a suitable choice for handling suprasegmental characteristics in speech including F0 and duration patterns. However, due to parallel training, applying seq2seq modeling to DVC results in unwanted change of speaker identity.

\subsection{Nonparallel frame-wise voice conversion based on variational autoencoder}

Nonparallel VC is attractive in that no parallel utterances between the source and target are required, and one of the major trends is autoencoder-based methods \cite{VAE-VC, VAE-GAN-VC, CHOU-NPVC, cyclevae, CDVAE-CLS-GAN}. Specifically, the encoder first encodes the input feature into a latent code, and the decoder then mixes the latent code and a specified target speaker embedding to generate the converted feature.
Autoencoders are usually trained with a reconstruction loss, but many techniques have been applied to solve various problems in training. The use of a variational autoencoder (VAE) \cite{VAE} is the most widely adopted method since it greatly stablizes training \cite{VAE-VC, VAE-GAN-VC, CDVAE-CLS-GAN}. Other techniques include using generative adversarial networks (GANs) to alleviate oversmoothing \cite{VAE-GAN-VC, CHOU-NPVC, CDVAE-CLS-GAN}, introducing a cyclic objective to improve conversion performance \cite{cyclevae}, or applying vector-quantization \cite{VQVAE} which introduces discreteness into the latent space to capture the categorial property of the linguistic contents in speech. 

VAE-based VC is categorized into the frame-based method, which tends not to convert supra-segmental features very well. Although the conversion similarity is therefore inferior to seq2seq-based methods, there are applications where it would be better to keep them unchanged, such as cross-linugal VC and, DVC.

\section{Proposed method}

An overview of our proposed method is illustrated in Figure~\ref{fig:method}. Assume that we have a parallel corpus between the dysarthria patient and multiple reference speakers. Our proposed method consists of two models: a seq2seq model that converts the acoustic feature sequence extracted from the input dysarthric speech into that of a reference normal speaker to be more intelligible and natural, and a nonparallel frame-wise model that restores the identity of the patient, which is realized by a VAE. To generate the converted waveform from the acoustic feature, we used the parallel waveGAN (PWG) neural vocoder \cite{parallel-wavegan} as the waveform synthesis module, which enables parallel, real-time waveform generation. Note that we do not perform waveform generation between the two models. In the following sections we explain details and design choices of the respective modules.

\subsection{Sequence-to-sequence modeling}
\label{ssec:seq2seq}

We adopted the many-to-many Voice Transformer Network (VTN) with text-to-speech (TTS) pretraining. VTN is a seq2seq model for VC based on the Transformer model \cite{transformer}, which relies on multi-head self-attention layers to efficiently capture local and global dependencies. It takes acoustic features (e.g. log mel spectrograms) as input and outputs converted acoustic features. It was extended to a many-to-many version in \cite{m2m-VTN}, which was shown to be more effective when a parallel corpus between multiple speakers is available.

The TTS pretraining technique is a two-stage process that transfers the core ability of a seq2seq VC model, which is to encode linguistic-rich hidden representations, from large-scale TTS datasets \cite{VTN, VTN-TASLP}.
First, the decoder pretraining essentially involves training a TTS model on a large-scale TTS dataset. Using the same TTS corpus as input and target, the encoder is then pretrained with a reconstruction loss by fixing the learned decoder from the first stage. Since the decoder was trained to recognize the linguistic-rich hidden representations encoded from text, the encoder hence learns to extract representations of similar properties from speech. The VC model training is finally performed with the VC corpus, which can be completely different from the TTS corpus in terms of speaker and content.

It is worth investigating the choice of the reference speaker. Although the corpus was designed to be parallel among the patient and all reference speakers, due to the difference in characteristics such as the speaking rate and F0 pattern, some speakers can be easier to convert to, compared to others. We thus hypothesize that by choosing a reference speaker with similar characteristics to the patient, conversion might be made easier. We define the similarity between the patient and the reference speaker to be the best performance the VC model can obtain. In later sections, we present our analysis on how the choice of reference speaker affects the conversion performance in various aspects.

\subsection{Nonparallel frame-wise model}

For the VAE model, we used \textit{crank} \cite{crank}, an open-source VC software that combines recent advances in autoencoder-based VC methods, including the use of hierarchical architectures, cyclic loss and adversarial training. To take full advantage of unsupervised learning, we trained the network using not only the data of the patient and the reference speakers but also a multi-speaker TTS dataset.

\section{Experimental Evaluations}
\label{sec:exp}

\subsection{Experimental settings}

To collect the dysarthric speech dataset, a female patient was asked to read the prompts in the phonetically-balanced TMHINT dataset \cite{tmhint}, where each of the 320 sentences contained 10 Mandarin characters. For the reference speakers, we used the audio recordings of 17 speakers (13 male and 4 female speakers)\footnote{Speaker SP11 was excluded due to labeling error.} in the TMSV dataset \cite{tmsv}, which was also collected with the TMHINT prompts. We used a 240/40/40 train/validation/test split. All speech utterances were downsampled to 16 kHz, and 80-dimensional mel spectrograms with a 16 ms frame shift were extracted as the acoustic feature.

The implementation of the VTN was based on the open-source toolkit ESPnet \cite{espnet, espnet-2020}. The detailed configuration can be found online\footnote{\url{https://github.com/espnet/espnet/tree/master/egs/arctic/vc1}}. The TTS pretraining was conducted with the Sinica COSPRO multi-speaker Mandarin dataset \cite{cospro}, which is 44 hr long. The implementation of VAE was based on \textit{crank}, which can also be accessed freely\footnote{\url{https://github.com/k2kobayashi/crank}}. Sinica COSPRO was used along with the TMSV and the patient's voice as training data for the VAE training. For the PWG, we followed an open-source implementation\footnote{\url{https://github.com/kan-bayashi/ParallelWaveGAN}}. The training data of PWG contained the audio recordings of the 18 TMSV speakers.

\subsection{Objective evaluation}

We carried out two types of objective evaluation. First, the mel cepstrum distortion (MCD) is a commonly used measure of spectral distortion in VC, which can only be calculated when the ground truth sample is available. We thus only used this metric in the evaluation of the VTN model. Second, to evaluate the intelligibility of the VC system, we used a Transformer-based automatic speech recognition (ASR) engine pretrained on the AISHELL-1 dataset \cite{aishell-1} to transcribe the converted speech, and directly calculated the character error rate (CER) based on the ASR outputs. We then converted the characters into pinyin and discarded the tone to obtain the syllable error rate (SER) of the converted speech.

%
\begin{figure}[t]
	\centering
	\includegraphics[width=\columnwidth]{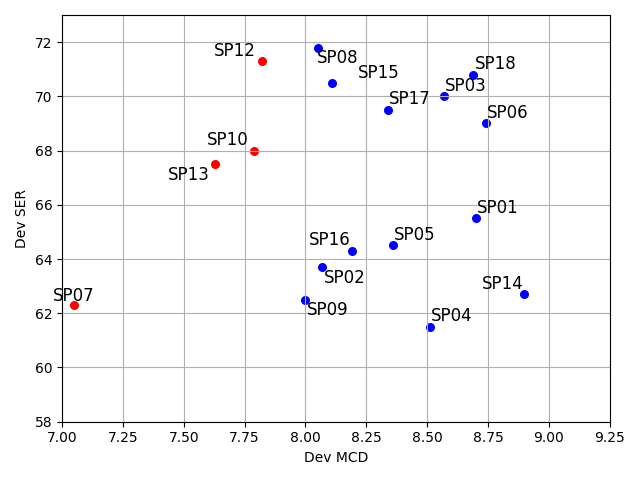}
	
	\caption{Scatter plots of the MCD and SER scores of each speaker. Both MCD and SER are the lower the better. Red and blue dots denote female and male speakers, respectively.}
	\label{fig:ref-spk-select}
	\vspace*{-2mm}
\end{figure}

\subsubsection{Investigation of the choice of reference speaker}

We first examine our hypothesis on the importance of the choice of reference speaker, as described in Section~\ref{ssec:seq2seq}. Since we carried out two types of objective metrics, it is worthwhile to examine which is a more proper selection criterion.
We trained the VTN model for 2000 epochs, and we selected the best performing models based on MCD. The results are shown in Figure~\ref{fig:ref-spk-select}.

First, for MCD, female reference speakers tend to yield lower scores, which is reasonable since the patient is also a female. On the other hand, the SER scores did not differ much between genders, and none of the genders gave obviously lower scores. Nonetheless, the speaker with the lowest MCD score (SP07) did not necessarily give the lowest SER value (SP09 gave the lowest SER value), and vice versa. To examine which criterion is better, we conducted a listening test as a more reliable proof, which will be presented in later sections.

\begin{figure}[t]
	\centering
	\includegraphics[width=\columnwidth]{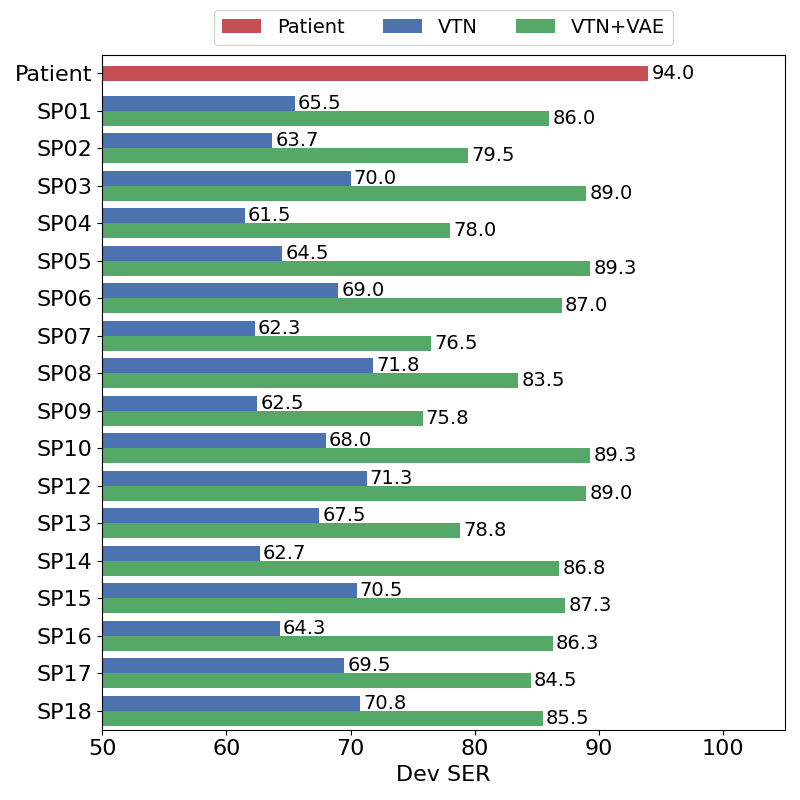}
	\centering
	\captionof{figure}{SER values of the patient's dysarthric voice and the output speech after the VTN model and the VAE model. The values are the smaller the better.}
	\label{fig:ser}
	\vspace*{-2mm}
\end{figure}

\subsubsection{Intelligibility degradation from VAE}
\label{sssec:obj-intel}

As stated in Section~\ref{sec:intro}, an important assumption in this work is that the intelligibility should be consistent throughout the VAE model. We examine how valid this assumption is by comparing the SERs of the original dysarthric voice as well as the output speech after the VTN model and the VAE model. The results are shown in Figure~\ref{fig:ser}. It can be clearly observed that our assumption was not held, as all SER values after the VAE model are much higher than those of the VTN model output. This is because of insufficient unsupervised factorization in the VAE model we used. As a result, a well-shared linguistic representations space between the normal speech and the dysarthric speech cannot be learned.

Nonetheless, the conversion pairs with most of the reference speakers still yielded lower values compared with the original dysarthric speech. Specifically, speaker SP09 gave the lowest SER of $75.8$ after the complete conversion process, which was $18.2$ points lower than the original $94.0$. This result demonstrates the effectiveness of the proposed two-stage method. In later sections, we will further examine the degradation of naturalness with the listening test results.

\begin{table*}[t]
	\centering
	\captionsetup{justification=centering}
	\caption{Results of subjective evaluation using the test set with 95\% confidence intervals. All values are higher the better.}
	\centering
	\begin{tabular}{ c c c c c c c c c }
		\toprule
		 & \multicolumn{4}{c}{Naturalness} & \multicolumn{4}{c}{Similarity} \\
		\cmidrule(lr){2-5} \cmidrule(lr){6-9}
		Description & SP04 & SP09 & SP07 & SP13 & SP04 & SP09 & SP07 & SP13 \\
		\midrule
		Dysarthric & \multicolumn{4}{c}{2.37 $\pm$ .19} & \multicolumn{4}{c}{\textemdash} \\
		TMSV & \multicolumn{4}{c}{4.99 $\pm$ .01} & \multicolumn{4}{c}{9\% $\pm$ 7\%} \\
		VTN & 3.29 $\pm$ .32 & 3.16 $\pm$ .27 & 3.45 $\pm$ .37 & 3.74 $\pm$ .27 & 8\% $\pm$ 8\% & 8\% $\pm$ 9\% & 30\% $\pm$ 11\% & 25\% $\pm$ 14\% \\
		VTN+VAE & 2.42 $\pm$ .30 & 2.38 $\pm$ .41 & 2.65 $\pm$ .39 & 2.60 $\pm$ .35 & 45\% $\pm$ 10\% & 45\% $\pm$ 14\% & 49\% $\pm$ 11\% & 42\% $\pm$ 11\% \\
		\bottomrule
	\end{tabular}
	\label{tab:sub-eval}
	\vspace*{-2mm}
\end{table*}

\subsection{Subjective evaluation}

We conducted subjective tests on naturalness and conversion similarity to evaluate the perceptual performance.
Since it is impractical to evaluate all converted samples of the 17 reference speakers, for both metrics we chose two speakers with the lowest values (MCD: SP07, SP13; SER: SP04, SP09). 
For naturalness, participants were asked to evaluate the naturalness of the speech by the mean opinion score (MOS) test on a five-point scale.
For conversion similarity, each listener was presented a natural target speech and a converted speech, and asked to judge whether they were produced by the same speaker on a four-point scale (Definitely the same, the same, different, definitely different). 
We recruited 11 native Mandarin speakers. Table~\ref{tab:sub-eval} shows the results.
Audio samples are available online\footnote{\url{https://bit.ly/3sHxaGY}}.

\subsubsection{Investigation of the choice of reference speaker}

We first continued our investigation on the reference speaker. For naturalness, as expected, the reference speakers with lower MCD values (SP07, SP13) outperformed the other two speakers (Sp04, SP09). Surprisingly, even after the VAE conversion, SP07 and SP13 still yielded better performances. This shows that listeners payed less attention to the intelligibility, but valued other factors such as fluency and stability more. This also explains why the dysarthric speech, although with extremely low intelligibility, still yielded a MOS score of $2.37$. On the other hand, for similarity, such trend was not so obvious, as only SP07 slightly outperformed the other two speakers, and the difference was not significant. Overall, the best performing reference speaker was SP07, whose naturalness ($2.65$) and similarity ($49\%$) scores were the best among all other speakers after VTN and VAE.

\subsubsection{Naturalness degradation from VAE}

Next, we continued to examine the naturalness consistency assumption described in Section~\ref{sec:intro}. it could be clearly observed that, regardless of which reference speaker, the naturalness scores degraded for almost 1 MOS point, showing that the current VAE model could not guarantee such consistency, which is similar to the findings in Section~\ref{sssec:obj-intel}. Nonetheless, the best performing speaker, SP07, yielded a naturalness MOS of $2.65$, which was shown to be significantly better than $2.37$, the MOS given by the dysarthric speech. This result again demonstrated the effectiveness of the proposed method.

\subsubsection{Identity preservation ability}

We finally examined the ability of our proposed method in maintaining speaker identity. Although the best similarity score of our method could achieve was only $49\%$, feedbacks from the listeners suggested that it was easy to find the converted speech different from that of the dysarthric speech due to its special characteristics. Since the normal speech of the patient is impossible to obtain, it is essentially difficult to evaluate conversion similarity. To this end, we concluded that the result was acceptable in this preliminary study, and would like to leave the improvement as future work.

\section{Conclusions and Discussions}

In this paper, we proposed a novel two-stage paradigm for maintaining speaker identity in DVC, where a parallel seq2seq model first converts the source dysarthric speech into that of a reference speaker with the quality enhanced, and a nonparallel frame-wise model realized by a VAE then converts the speaker identity back to the patient while preserving the quality. The experimental results showed that (1) the reference speaker with lower MCD is considered better, (2) the current VAE model does not guarantee quality consistency, and (3) our method can still improve the quality to a certain extent while preserving speaker identity. Yet, the current performance is still far from satisfactory, and below we discuss several improving directions.

\noindent{\textbf{Improve seq2seq modeling.}} The current intelligibility after the first seq2seq conversion stage was much worse than that of past DVC works on simpler datasets \cite{dysarthric-e2e}. Although we believe this is due to the more complex, limited dataset we used, it is worthwhile to apply techniques like text supervision \cite{S2S-Text-VC} or data augmentation \cite{S2S-parrotron-VC}.

\noindent{\textbf{Quality consistency assumption.}} The current VAE model we employed could not guarantee to preserve the enhanced quality. Possible directions include a hierarchical structure to modify solely time-invariant characteristics, or resorting to other frame-wise models such as PPG-based methods \cite{VC-PPG}.

\noindent{\textbf{Automatic reference speaker selection.}} In this work, we chose the best reference speaker by examining the MCD and SER scores of the converted speech, which was an ad-hoc approach. To quickly decide the suitable reference speaker for an arbitrary patient, we may further use pretrained speaker representations like x-vectors \cite{x-vector} as a measurement.

\section{Acknowledgements}

We would like to thank Dr. Hirokazu Kameoka from NTT CS Laboratory, Japan for the fruitful discussions. This work was supported by JST CREST Grant Number JPMJCR19A3, Japan.

\bibliographystyle{IEEEtran}

\bibliography{ref}
\end{document}